\documentclass[12pt,a4paper,twoside]{article}
\usepackage{epsfig}
\usepackage{baltlat1}
\usepackage{wrapfig}
\begin{document}
\ \
\vspace{0.5mm}

\setcounter{page}{1}
\vspace{8mm}

\titlehead{Baltic Astronomy, vol.13, 193--200, 2004.}

\titleb{THE BEPPOSAX REVOLUTION IN GAMMA--RAY \\ BURST SCIENCE}

\begin{authorl}
\authorb{L.~Amati}{}
\end{authorl}

\begin{addressl}
\addressb{}{Istituto di Astrofisica Spaziale e Fisica Cosmica - Sez. Bologna,
CNR, via P. Gobetti 101, I-40129 Bologna, Italy}
\end{addressl}

\submitb{Received October 15, 2003}

\begin{abstract}
We review the BeppoSAX discoveries and studies in the field of
Gamma-Ray Bursts, which allowed a huge step forward in the understanding of
these still mysterious and highly interesting astrophysical sources.
\end{abstract}


\resthead{The BeppoSAX revolution in GRB science}{L.~Amati}

%{Institution}{Author(s)}

%\def\ninepoint{\def\rm{\fam0\ninerm} \textfont0=\ninerm}

\sectionb{1}{THE BEPPOSAX MISSION AND ITS SCIENTIFIC PAYLOAD}

BeppoSAX was a mission of the Italian Space Agency (ASI) with contribution of 
NIVR (Netherlands)
launched in April 1996 and ended in April 2002. It was characterized by a
Low Equatorial Orbit with 3.9$^{\circ}$ inclination and $\sim$600 km altitude, 
resulting in a low and very stable background (few \% variation along a single orbit). 
The scientific payload covered an unprecedented broad energy band (0.1--300 keV) and
was composed by two set of instruments: the Narrow Field Instruments (NFI),
pointing along the satellite $Z$ axis, and two wide field instruments pointing
in directions perpendicular to it. The NFI included (we report in parenthesis
the energy band, the FOV and the angular resolution): the LECS (0.1--10 keV, 
0.5$^{\circ}$, $\sim$3'), the MECS, composed by three units,
(1.3--10 keV, 0.5$^{\circ}$, $\sim$1'), the HPGSPC (4--120 keV, 1.1$^{\circ}$,
collimated) and the PDS (15--300 keV, 1.3$^{\circ}$, collimated).
The wide field instruments were the Wide Field Cameras (WFC), two units pointing along
the $\pm$X axis (2--28 keV, 20$^{\circ}$ FOV, 3' ang. res.), and the Gamma--Ray Burst Monitor,
four units pointing along the $\pm$X and $\pm$Y directions (40--700 keV, open FOV,
burst location capabilities of several degrees).
See Boella et al. (1997) and references therein for more details.

\sectionb{2}{THE DISCOVERY OF GAMMA-RAY BURSTS AFTERGLOW}

%\subsectionb{2.1}{GRB~960720: the first few arcmin localized event}
The possibility of detecting GRBs simultaneously with the GRBM and the WFC is implicit
in the payload design described above and was 
already considered in the BeppoSAX Observers Handbook, 
with $\sim$6 events/year expected.
The first of such events was GRB~960720, discovered in the WFC and localized within 
3' with off-line analysis. A Target Of Opportunity (TOO) observation with
the NFI was performed 45 days after the GRBM detection, but 
no convincing counterpart was found (Piro et al. 1998). 
Nevertheless, this result gave great confidence 
on the BeppoSAX capabilities of accurately localizing GRBs and
an on-line procedure was designed and activated at the Satellite Operation Center (SOC)
in order to allow for prompt (within 1 orbit duration, 90 minutes) search and localization of GRBs simultaneously detected by the GRBM and WFC.
Soon after, this procedure, which exploited at the best the BeppoSAX capabilities, led 
to the discovery of the first X-ray afterglow source of a GRB (GRB~970228), with the NFI 
(Costa et al. 1997). The accurate localization of the X--ray afterglow by the NFI allowed
the discovery of an optical transient associated to this event, the first optical
counterpart to a GRB (Van Paradijs et al. 1997). Subsequently, the accurate localization, 
fast follow--up and afterglow discovery of another event, GRB~970508, led to the
measurement of the first redshift (cosmological !) of a GRB (Metzger et al. 1997) and to the
discovery of the first radio afterglow source of a GRB (Frail et al. 1997), allowing the measurement of its compactness and highly relativistic expansion.
Of the 51 GRBs simultaneously detected by the BeppoSAX GRBM and WFC and promptly alerted,  37 were followed up with the BeppoSAX NFI.
X--ray afterglows were discovered for $\sim$90\% of the followed-on GRBs, vs. 
$\sim$50\% in the optical and $\sim$40\% in the radio. Redshift estimates were obtained for 14 of these events. 
All events detected with WFC and GRBM are long ($>$2 s), but the GRBM detected also short GRBs.\\
These unprecedented discoveries settled the 30 year old  question about GRBs distance, 
allowed first strict testing  of GRB models, provided a
new phenomenology and opened new observational windows for the study of these sources.

\sectionb{3}{PHYSICS OF GAMMA--RAY BURSTS}

\subsectionb{3.1}{Study of the prompt emission from X to gamma--rays}

The co--alignment of the WFC units with two of the GRBM detectors allowed not only the
accurate localization of those GRBs detected by both instruments but also the study of
their prompt emission from $\sim$700 keV down to 2 keV with unprecedented accuracy and good 
calibration. The comparison between the X--ray and gamma--ray light curves allowed the study
of pulse width $\Delta t$ as a function of photon energy $E$, confirming in 
several cases
the synchrotron predictions of a power--law dependence $\Delta t \propto E^{-0.5}$
(e.g. Piro et al. 1998, Feroci et al. 2001). The extension
to low energies of the spectral measurements allowed a more accurate estimate of the low energy spectral index (a crucial observable for prompt emission models), and its evolution, 
with respect e.g. to BATSE ($>$25 keV).
Most ($\sim$70\%) of the time averaged spectra of GRBs detected with BeppoSAX are well 
fit down to 2 keV with an optically thin synchrotron shock model (Tavani 1996).
The spectral evolution from X to gamma--rays shows 
in general a hard-to-soft evolution, which may be due to 
decrease of the magnetic field in the post-shock region due to its expansion 
or post-shock decrease of the electron energy distribution due to cooling processes.
The X--ray spectra accumulated during the first part of the prompt emission are sometimes inconsistent with optically thin synchrotron,
indicating that another mechanism, likely Inverse Compton, may be at work at early times. 
See Frontera et al. (2000a) for details.

\subsectionb{3.2}{X--ray afterglow}

The X--ray afterglow of GRBs are characterized by a power-law decay and a  
power-law spectrum. The distribution of the decay indices $\delta$ is a Gaussian with 
$<\delta>$ =1.33 and $\sigma$= 0.33.
The distribution of the spectral photon indices $\Gamma$ is a Gaussian too, with  
$<\Gamma>$ = 1.93 and  $\sigma$ = 0.35. These findings are in agreement with the
predictions of external shocks
synchrotron models (e.g. Sari et al. 1999), also if in some cases (e.g. GRB~970508, 
see Piro et al. 1999) deviations from the monotonic decay have been observed. 
Also multi-wavelength spectra, combining the X--ray data with those at lower wavelengths, are crucial to establish the true mechanism(s) at work. 
Consistency with synchrotron predictions was found e.g. for GRB~970508 (Galama et al. 
1997), but 
deviations from the predicted spectral shape in the X-ray band were found for GRB~000926 (Harrison et al. 2001) and GRB~010222 (in 't Zand et al. 2001).
These deviations are consistent with relevant Inverse Compton on the low energy 
photons.

\subsectionb{3.3}{Connection between prompt and afterglow emission}

Evidence that the X-ray afterglow starts during the tail of the prompt 
emission emerged from the BeppoSAX data.
In several cases (e.g., GRB~970228) this evidence is direct, because the extrapolation
of the power--law best fitting the 2--10 keV afterglow light curve is consistent
with the 2--10 keV flux measured by the WFC during the last part of the prompt emission.
A study of a sample of GRBs simultaneously detected by the WFC and GRBM (Frontera et al. 2000a) showed that: a) prompt emission spectra evolve toward the spectra of the late 
afterglow; b) the 2-10 keV fluence of the late prompt emission and its evolution 
are
consistent with that of the afterglow emission.
The afterglow onset time inferred from the later analysis was found to occurr at 
$\sim$60\% of 
the
GRB duration, on average. This allowed the estimate of the
initial Lorentz factor 
of the shocked fluid, with a mean value found of 150. 

\subsectionb{3.4}{Breaks in the afterglow light curves: indications of jet}

In  case of a jet-like geometry of the GRB emission, an achromatic break in the 
afterglow light curve is expected at the time when the beaming angle starts to exceed 
the jet angle.
Breaks in the light curve may also be due to transition to non--relativistic (NRE) 
phase of the expanding fireball.
Several detections of breaks in the optical and radio bands have been reported. 
Indication of a break in the GRB~990510 X-ray afterglow,
consistent with that observed in the optical band 
and with a theoretical model, was found by Pian et al. (2001) by comparing the
afterglow data with those of the prompt emission.
In the same way, a break in the GRB~010222  X-ray afterglow fading law was
found by in't Zand et al. (2001), but in this case the 
break is consistent with a transition to a NRE phase.
Our systematic analysis of all the BeppoSAX GRBs for which there was detection 
of X--ray 
afterglow confirms these results. The post-break temporal indices derived from 
X-ray light curves, when plotted vs. their
spectral photon indices, show that two GRBs (990510 and 010214) are consistent 
(basing on the relationships predicted by Sari et al. 1999) 
with the emission from a spreading jet, while the third (GRB~010222) cannot be 
distinguished by a an isotropic expansion.
The fact that at least some GRBs seem to be collimated and the study of the 
distribution of the
jet angles have a strong impact on the understanding of the energetics and the
progenitors of these phenomena.

\sectionb{4}{ABSORPTION AND EMISSION FEATURES}

\subsectionb{4.1}{Transient absorption features in the early prompt emission}

A transient absorption edge at $\sim$3.8 keV in the first 13 s of the prompt
X--ray emission of GRB~990705, associated with a variable column density $N_H$,
was discovered by Amati et al. (2000).
This feature was interpreted as a redshifted
K absorption edge of neutral Fe within a shell of material around the GRB site, 
photo-ionized by the GRB photons. The consequences of this interpretation were:
estimate of the X--ray redshift (0.86$\pm$0.17) of the burst source, which  was later 
confirmed by the optical redshift of the GRB host galaxy (Le Floch et al. 2002);
an iron abundance relative to solar of $\sim$75, typical of supernova explosions;
a large mass of Fe, unless Fe is clumped and a clump is along the line of sight.
An alternative interpretation, allowing to reduce the implied iron relative abundance 
and mass, is that the absorption line is due to resonant scattering of GRB photons off 
H--like Fe (transition 1s-2p, $E_{rest}$ = 6.927 keV) (Lazzati et al. 2001). 
Variable intrinsic absorption was also found in the prompt emission
of GRB~980329. 
The $N_H$ time behavior of this source can be explained e.g. if the GRB 
event occurs in  over-dense regions within  molecular clouds (Bok globules).
A very recent spectral analysis performed on the whole sample of GRBs simultaneously
detected by the WFC and GRBM put in evidence the presence of a transient absorption
feature in the X--ray spectrum of GRB~011211 and of a variable column density in 
the X--ray spectrum of GRB~000528 (papers in preparation).

\subsectionb{4.2}{X-ray lines in the afterglow spectra}

BeppoSAX discovered X-ray emission features in the afterglow spectra of 
GRB~970508, first case ever ($E_l$ =3.4$\pm$0.3 keV, Piro et al. 1999), and
GRB~000214 ($E_l$ =4.7$\pm$0.2 keV ) (Antonelli et  
al. 2000). 
This lines were interpreted as redshifted neutral or ionized Fe lines.
X-ray emission features were discovered also by other missions (ASCA, Chandra, 
XMM-Newton), e.g. in the afterglow spectra of
GRB~970828,
%($E_l$ =$\sim$5.1 keV) (Yoshida et al. 1999),
GRB~991216 and
%($E_{l1}$ =3.49$\pm$0.06 keV; $E_{l2}$ =4.4$\pm$0.5 keV)  
%(Piro et et al. 2000),
GRB~011211.
% (evidence of 5 lines between 0.44 and 1.46keV) (Reeves et al. 2002).
Very important information on the circum--burst material and the nature of the
progenitor were inferred from the observed X-ray lines: 
high density medium surrounding the GRB;
overabundance of metals (up to $\sim$60 times the solar abundance);
high velocity outflow of the X-ray emitting plasma (up to $\sim$0.1c).
X-ray lines rule out the NS merger models and strongly point to an environment typical 
of a young supernova explosion.

\sectionb{5}{OTHER TOPICS AND MORE RECENT RESULTS}

\subsectionb{5.1}{The GRB/SN connection}

One of the two candidate sources (S1, Pian et al. 2000) of the afterglow emission 
from GRB~980425 was coincident with SN1998bw in ESO 184-G82 (z=0.0085).
Likely S2 was a source field and  S1 was the afterglow source, but it showed a fading 
with unusual power-law index ($\delta$ =0.2). 
The unusual behavior was confirmed by the 6 month later BeppoSAX  observation.
The association of GRB 980425 with Type Ic SN 1998bw is now established.
Likely the detected X-ray light curve is the superposition of two components: 
a) the GRB afterglow continuum emission;
b) the SN emission from a peculiar type Ic SN.
These results were the first observational support to the GRB/SN connection hypothesis,
reinforced by subsequent indications of bumps in the light curves of some GRBs and
very recently by the evidence of a connection between
GRB~030329 and SN2003 (Stanek et al. 2003, Hjorth et al. 2003).

\subsectionb{5.2}{Dark and X--ray rich GRBs }

About 50\% of the BeppoSAX accurately localized and promptly followed-up GRBs do not 
have an optical counterpart, and are therefore classified as dark GRBs.
Statistical studies show that the optical searches of these events have been carried 
out to magnitude limits fainter on average than the known sample of optical afterglow.
Some of the dark GRBs could be intrinsically faint events, but this fraction cannot be 
high because the majority of them show X-ray afterglow emission comparable to that 
observed in GRBs with optical afterglows
The possible explanations of dark GRBs include: very high z ($>$5, with optical flux 
absorbed by intervening Ly$\alpha$ forest clouds), very dense circum--burst material, 
obscuration by scattering with dust in their host galaxy.
In addition to dark GRBs, there are 3 GRBs (981226, 990704 and 000615) belonging to 
the WFC/GRBM sample that show a ratio 
between the X-ray and gamma-ray fluence (or peak flux) much higher than the average.
These X-ray rich GRBs are dark and their X--ray afterglows do not show peculiar
characteristics, even if the decay index of GRB~990704 (Feroci et al. 2001) 
is quiet flat and the
afterglow of GRB~981226 show indication of a rise during the beginning of the first 
BeppoSAX TOO (Frontera et al. 2000b). 
Among possible explanations of X--ray rich GRBs there are a low Lorentz factor, due 
e.g. to 
high barion loading of the fireball, or a very high $z$. 
 
\subsectionb{5.3}{X--Ray Flashes}
 
Among the FXTs detected by the BeppoSAX WFC there are $\sim$30 events showing the same 
properties of the X-ray counterparts of GRBs (light curve shape, duration, non thermal 
spectrum, isotropic distribution) but with no corresponding signal in the GRBM: 
new events called X-Ray Flashes (XRFs).
In principle, XRFs could be a different class of sources, 
but there is evidence that they
 are a sub-class of GRBs: a) for most events the extrapolation of WFC spectra to high 
energies are consistent with GRBM upper limits; b) an off--line inspection of BATSE 
light curves in 25-100 keV shows that 9 out of 10 XRFs observable by this instrument 
were actually detected (Kippen et al. 2003); c) very recently, X-ray afterglow emission 
was detected 
for three XRFs (011030, 020427, 030723), see e.g. Amati et al. (in preparation).
XRFs extend to very low energies the distribution of the spectral peak energies of 
GRBs, which is thus much 
broader than inferred from  CGRO/BATSE (narrow distribution around 200 keV).
Possible interpretations of the nature of these sources include: events with very low 
Lorentz factor (due e.g. to high  barion loading of the fireball), events at very high 
z, collimated events seen very off-axis.

\newpage
\subsectionb{5.4}{The $E_p^{rest}$ vs. $E_{rad}$ relationship}

Basing on a sample of BeppoSAX GRBs with known redshift, Amati et al. (2002) found
that the peak energy of the intrinsic EF(E) spectrum, $E_p^{rest}$, is correlated,
with a power--law dependence with index $\sim$0.5, to the isotropic equivalent radiated
energy $E_{rad}$.
Such dependence, confirmed by HETE--2 measurements and possibly holding also for XRFs
(Lamb et al. 2004), could be explained  if 
the radiation is due to synchrotron and
it is isotropically emitted or 
it is emitted with the same beam angle. 
In any case, it can put strong constraint to the models for the prompt emission
of GRBs (see e.g. Zhang \& Meszaros 2002).

\References
\def\ref{\vskip.3mm
\hangindent12pt\hangafter=1
\noindent\ignorespaces}
\ref
Amati~L., Frontera~F., Vietri~M. et al.  2000, Science, 293, 953
\ref
Amati~L., Frontera~F., Tavani~M. et al.  2002, A\&A, 390, 81
\ref
Antonelli~L.A., Piro~L., Vietri~M. et al. 2000, ApJ, 545, L39
\ref
Boella~G., Butler~R.C., Perola~G.C. et al. 1997, A\&AS, 122, 327
\ref
Costa~E., Frontera~F., Heise~J. et al. 1997, Nature, 387, 783
\ref
Feroci~M., Antonelli~L.A., Soffitta~P. et al. 2001, A\&A, 378, 441
\ref
Frail~D., Kulkarni~S.R., Costa~E. et al. 1997, IAU Circ. 6576
\ref
Frontera~F., Amati~L., Costa~E. et al. 2000a, ApJS, 127, 59
\ref
Frontera~F., Antonelli~L.A., Amati~L. et al. 2000b, ApJ, 540, 697
\ref
Galama~T.J, Wijers~R.A.M.J., Bremer~M. et al. 1997, ApJ, 500, L97
\ref
Harrison~F.A., Host~S.A., Sari~R. et al. 2001, ApJ, 559, 123
\ref
Hjorth J., Sollerman J., Moller P. et al. 2003, Nature, 423, 847
\ref
Kippen~R.M., Woods~P.M., Heise~J. et al. 2003, AIP Conf. Proc., 662, 244
\ref
Lamb~D.Q., Donaghy~T.Q., Graziani~C. 2004, NewAR, 48, L429
\ref
Lazzati~D., Ghisellini~G., Amati~L. et al. 2001, ApJ, 556, 471
\ref
Le Floc'h~E., Duc~P.--A., Mirabel~I.F. et al. 2002, ApJ, 581, L81
\ref
Metzger~M.R., Djorgovski~S.G., Kulkarni~S.R. et al. 1997, Nature, 387, 878
\ref
Pian~L., Amati~L., Antonelli~L.A. et al. 2000, ApJ, 536, 778
\ref
Pian~L., Soffitta~P., Alessi~A. et al. 2001, A\&A, 372, 456
\ref
Piro~L., Heise~J., Jager~R. et al. 1998, A\&A, 329, 906
\ref
Piro~L., Costa~E., Feroci~M. et al. 1999, ApJ, 514, L73
\ref
Sari~R., Piran~T., Halpern~J.P. 1999, ApJ, 519, L17
\ref
Stanek~K.Z., Matheson~T., Garnavich~P.M. et al. 2003, ApJ, 591, L17
\ref
Tavani~M. 1996, ApJ, 466, 768
\ref
Van Paradijs~J., Groot~P.J., Galama~T. et al. 1997, Nature, 386, 686
\ref
in 't Zand~J.J.M., Kuiper~L., Amati~L. et al. 2001, ApJ, 559, 710
\ref
Zhang~B., Meszaros~P. 2002, ApJ, 581, 1236

\end{document}